\documentclass[12pt,letterpaper]{JHEP3}
\usepackage{amsmath,amssymb,scalefnt}
\usepackage[dvips]{epsfig}

\newcommand{\abbrev}{\scalefont{.9}}
\newcommand{\susy}{{\abbrev SUSY}}
\newcommand{\sm}{{\abbrev SM}}
\newcommand{\gut}{{\abbrev GUT}}
\newcommand{\guts}{{\abbrev GUTS}}
\newcommand{\mssm}{{\abbrev MSSM}}
\newcommand{\nmssm}{{\abbrev NMSSM}}
\newcommand{\mnmssm}{{\abbrev MNMSSM}}
\newcommand{\sugra}{{\abbrev SUGRA}}

\def\phidot{\dot\phi}
\def\phiddot{\ddot\phi}
\def\Hdot{\dot H}
\def\sic{supersymmetric}

\def\be{\begin{equation}}
\def\ee{\end{equation}}
\def\bea{\begin{eqnarray}}
\def\eea{\end{eqnarray}}

\def\GeV{\hbox{GeV}}
\def\TeV{\hbox{TeV}}

\def\nn{\nonumber\\}
\def\pa{\partial}

\def\Dbar{{\overline{\cal D}}}

\author{Martin B Einhorn$^{1}$, D R Timothy Jones$^{1,2}$\\
$^{1}$Kavli Institute for Theoretical Physics, 
University of California, Santa Barbara
CA 93106-4030, USA\\
$^{2}$Dept. of Mathematical Sciences,
University of Liverpool, Liverpool L69 3BX, UK}
\title{Inflation with Non-minimal Gravitational Couplings and Supergravity}
\abstract{We explore in the supergravity context the possibility that a Higgs
scalar may  drive inflation via a non-minimal coupling to gravity 
characterised by a large dimensionless coupling constant.   
We find that this scenario is not compatible with the \mssm, but that adding a singlet field 
(\nmssm,  or a variant thereof) can very naturally 
give rise to slow-roll inflation. The inflaton is necessarily contained 
in the doublet Higgs sector and occurs in the $D$-flat direction of the two Higgs doublets.}
\preprint{LTH 860\\NSF-KITP-09-216\\}
\begin{document}


\section{Introduction}

The idea that a large non-minimal coupling of scalar fields to gravity 
might play an important role in inflation is clearly worth exploring. 
(For an early example see Ref.~\cite{Salopek:1988qh}).  In particular,
there has been recent
interest~\cite{Bezrukov:2007ep}-\cite{Okada:2009wz} in the economical
possibility that the standard model (\sm) Higgs $H$ can be  the relevant
scalar; the action $S$ is  \be  S = \int\, d^4 x \sqrt{-g} \left[ M_P^2
R -\xi R H^{\dagger} H + {\cal L}_{SM}+ \cdots \right] \label{eq:bs} \ee
where $R$ is the Ricci scalar, and $\xi$ is a dimensionless coupling
constant, assumed positive, \footnote{Here, the action has been
expressed in the Jordan frame; one may transform to the Einstein frame
by a conformal transformation.} and ${\cal L}_{SM}$ is the Lagrangian of
the SM,  generalized to a nontrivial background spacetime obtained by
replacing the  flat-space metric $\eta_{\mu\nu}$ with the curved space
metric $g_{\mu\nu}.$     Classically, for large enough values of $\xi$
and $h$, viz.,  \be \xi h^2 \gtrsim M_P^2 \gg h^2, \label{eq:xibiga} \ee
(where $H = (0,h)^T$), it transpires that the scalar potential is nearly
flat and the standard slow-roll approximation is possible.   The issue
of whether this classical analysis remains valid as an effective quantum
field theory for field values in the range Eq.~(\ref{eq:xibiga}) is a
matter of debate at present.   On the one hand, there are
some~\cite{Burgess:2009ea,Barbon:2009ya} who argue that, as an effective
field theory, the full action Eq.~(\ref{eq:bs}) must include higher
dimensional operators, such as $\xi^2 (H^{\dagger} H)^6/M_P^2,$ so that
one would expect the theory to work up to an  energy scale
$\Lambda=M_P/\xi,$ which is far below $\Lambda=M_P/\sqrt{\xi},$ the
regime suggested by Eq.~(\ref{eq:xibiga}).  

On the other hand, others
argue the effective field theory works within the regime in
Eq.~(\ref{eq:xibiga}) for a variety of reasons.  One such
argument~\cite{Salopek:1988qh, DeSimone:2008ei, Bezrukov:2009db,Barvinsky:2009fy} is
that loops involving virtual Higgs are suppressed because their
propagators are modified from ordinary perturbation theory for large
values of the Higgs field.  The treatment of gravity as a classical
background may also be challenged.  One would normally expect that the
feedback on the metric would become strong on scales of the order of
$M_P,$ but one may hope~\cite{Salopek:1988qh, Bezrukov:2007ep} that, in
some models, the the energy density associated with Higgs remains small
compared to $M_P^4,$ even for $h>O(M_P),$ provided it is not
very much larger, say $h\lesssim O(10M_P).$  On the other hand,
Higgs-graviton scattering would seem to become strong or to violate
unitarity for energies $E\sim M_P/\xi$~\cite{Burgess:2009ea,Barbon:2009ya}.  
We will not address this issue directly in this paper, but we will return to it 
again in our conclusions.

The occurrence of such non-minimal couplings of matter fields to the
background curvature is a fact of life in the SM, because the notion of
minimal coupling ($\xi=0$) is only valid classically.   In quantum field
theory, the renormalized $\xi$ runs, i.e, is a function of a scale
parameter ($\xi=\xi(\mu)$) for which  $\xi=0$ is not a fixed
point~\cite{Jack:1984vj,Buchbinder:1992rb}.  Thus, the ``minimal
theory'' is at best an approximation to cases where the contributions of
terms like the $\xi$ term above may be neglected.

Whether or not perturbation theory remains viable in the scenario, the
scales of interest are large compared to the electroweak unification
scale, so that  the \sm{} becomes unnatural, suffering also other
apparent shortcomings such as unsatisfactory grand unification and lack
of a suitable Dark Matter candidate.  The latter issue is addressed, for
example, in Refs.~\cite{Clark:2009dc,Lerner:2009xg} by adding a singlet
to the standard model.  More  attractive alternatives to the SM above the
\TeV-scale are supersymmetric extensions of \sm{}, which overcome the
three most glaring deficiencies of the SM mentioned.  An obvious
question is whether the inflationary scenario described above survives in
such supersymmetric models, and, if so, what  their
properties are.  In this paper, we shall consider the \mssm{} and a
simple extension with an additional singlet field (\nmssm), with which
we will assume the reader is familiar (for a review see
Ref.~\cite{Ellwanger:2009dp}). 

An outline of the remainder of this paper is as follows: in the next
section, we review the standard form of supergravity and then consider
the addition of a non-minimal interaction analogous to the above.  In
Section~3, we apply this formalism to the \mssm{} , following in
Section~4 with the \nmssm{}.  Section~5 contains an analysis of  the
slow roll parameters in the one successful scenario that we identified. 
In Section~6, we restate our conclusions and discuss issues for future
work, commenting on some dramatic differences from the SM to be
anticipated for radiative corrections resulting from supersymmetry.

\section{Non-minimal couplings in supergravity}

Although particle physics generally ignores gravity in generalizing to
supersymmetry, its inclusion is mandatory in the present context.  Thus,
the natural starting point is not global supersymmetry but the 
modification of the Lagrangian for supergravity\footnote{Our notation
and conventions  follow Wess and Bagger~\cite{Wess:1992cp}, and from now
on we set $M_P = 1$.} coupled to a multiplet of chiral superfields
$\Phi$:

\be
{\cal L} = -6 \int\, d^2\Theta\, {\cal E} \left[ R - \frac{1}{4}(\Dbar^2 - 8R)\Phi^{\dagger}\Phi + P(\Phi)\right] 
+ h.c. + {\rm gauge~terms}.
\label{eq:sugra}
\ee
where ${\cal E}$ is the vierbein multiplet, $R$ is the curvature
multiplet containing the scalar curvature in its $\Theta^2$  component, and
$P(\Phi)$ is the superpotential.  For our purposes, we will not need the
explicit form of the gauge field contributions.

To extend Eq.~(\ref{eq:bs}) to the \sic\ case we replace
Eq.~(\ref{eq:sugra}) by  
\be {\cal L}_\chi = -6 \int\, d^2\Theta\, {\cal E}
\left[ R + X(\Phi) R - \frac{1}{4}(\Dbar^2 - 8R)\Phi^{\dagger}\Phi +
P(\Phi)  \right]  + h.c.  + {\rm gauge~terms}
\label{eq:sugrabs} 
\ee 
where $X(\Phi)$ is
quadratic in $\Phi$, with coefficients which are {\it dimensionless\/}
coupling constants.  We shall discuss examples below, but an immediate
consequence of \susy\ is that  the gauge invariant function $X(\Phi)$
is a function of the chiral superfields $\Phi$ only and not $\Phi^\dagger.$  
A second consequence is that, by the non-renormalization theorem, the various monomials in $X(\Phi)$ will 
(in a similar manner to those in the superpotential $P(\Phi)$) {\it not\/} be generated through radiative corrections 
if absent from the classical action  (in contrast to the case of the \sm{} $\xi$ from Eq.~(\ref{eq:bs})).

The scalar potential in the Einstein frame for the theory defined by Eq.~(\ref{eq:sugra})  is given by 
\bea
V(\phi,\phi^*) &=&  e^G\left[G_i (G^{-1})_{i j^*} G_{j^*}- 3\right]+ V_D(\phi,\phi^*)\nn
&=&  e^K\left[(K^{-1})_{i j^*}D_i P(D_j P)^* - 3P P^*\right] + V_D(\phi,\phi^*)
\label{eq:Vpot}
\eea
where $G = K + \ln (PP^*)$, $K(\phi,\phi^*)$ is the K\"ahler potential, 
and  the $D$-terms $V_D(\phi,\phi^*)$ will be  discussed below.  
We will take $K$ to have the canonical form 
\be
K = -3\ln (-\Omega/3),
\label{eq:ka}
\ee
where
\be
\Omega = \phi^{*}_i\phi_i - 3.
\label{eq:om}\ee
In Eq.~(\ref{eq:Vpot}), $D_i P = \pa_i P + K_i P,$ etc.

One can show that the effect  of the non-minimal couplings on the scalar potential, generalising from 
Eq.~(\ref{eq:sugra}) to Eq.~(\ref{eq:sugrabs}),
is simply to replace $\Omega$ in Eq.~(\ref{eq:om}) by 
\be
\Omega_{\chi} = \phi^{*}_i\phi_i - 3 - \frac{3}{2}(X(\phi) + h.c.).
\label{eq:omxi}
\ee

Let us turn to the $D$-term. In general, this takes the form (for a simple group with gauge coupling $g$)
\be
V_D = \frac{g^2}{2}\hbox{Re} f^{-1}_{ab} G_i (T^a)_{ij}\phi_j G_k (T^b)_{kl}\phi_l,
\ee
where $f_{ab}$ is associated with the kinetic energy of the gauge field
and, therefore, must  be a holomorphic function of $\phi_i$, and where
$\phi_i$ transform according to a representation $T^a$ of the gauge
group.  Noting that
\be
\pa_i P  (T^a)_{ij}\phi_j = \pa_i X (T^a)_{ij}\phi_j = 0
\ee
by gauge invariance, and assuming for simplicity the canonical form $f_{ab} = \delta_{ab}$, we find 
\be 
V_D = \frac{g^2}{2}\frac{9}{\Omega_{\chi}^2}(\phi^* T^a \phi)^2
\ee
Thus in the \mssm, the $D{\rm -terms}$ 
are 
\be
V_D = \frac{9}{\Omega_{\chi}^2}\left[\frac{g'^2}{8}\left( H_1^\dagger H_1- H_2^\dagger H_2 \right)^2+ \frac{g^2}{8}\left( \sum_{i=1,2} 
H_i^\dagger \vec{\tau} H_i \right)^2\right],
\label{eq:Vd}
\ee
where $H_1$ and $H_2$ are the two Higgs doublets, and $ \vec{\tau}$ are the Pauli matrices.

The behavior of the potential depends in detail on the choice of
superpotential $P(\phi)$, but we will be interested in directions in
field space in which the potential is approximately constant for large
values of the Higgs fields.  

\section{\mssm}

Let us begin by considering the effect of Eq.~(\ref{eq:omxi}) on the
scalar potential for the \mssm. The unique possibility for $X$ in this case 
is
\be
X =  \chi H_1 H_2,
\ee
where $\chi$ is constant; and, dropping terms involving fields that do not appear in $X$, 
the most general possibility for the superpotential $P$
is
\be
P = \Lambda + \mu H_1 H_2,
\ee
where $\Lambda$ and $\mu$ are constants. So $P$ contains no dimensionless couplings. 
One cannot in general choose both $\mu$ and $\chi$ to be real; 
however,  we will ignore this 
potential source of $CP$-violation in what follows, since, as we shall soon see, this case does not yield a 
suitable inflationary regime.   Without loss of generality, we may choose $\chi>0.$  

Let us calculate the potential in the electromagnetism-preserving direction
\be
H_1 = (h_1,0)^T, H_2 = (0, h_2)^T, 
\label{eq:dflat}
\ee 
where we may assume $h_{1,2}$ are real.   
If, as usual, we define   $h_1 = h\cos\beta$ and $h_2 = h \sin\beta,$
we can then write 
\be
P=\Lambda +\mu h^2\sin(2\beta)/2, ~~~~~~~~~ X=\chi  h^2\sin(2\beta)/2.
\ee
We shall be interested in the case when $\chi\gg 1,$ but $h^2\ll 1,$  so we must have  
$\chi\sin(2\beta)\ge0$ to avoid a singularity in $\Omega_\chi$.  Since $\chi>0,$ we 
must take $0\le\beta\le\pi/2 (\rm{mod}~\pi).$
We find  
\bea
V = -12\ \frac{(6 \Lambda+\mu h^2\sin 2\beta)^2 +12\chi\mu h^2 ( 3 \Lambda+\mu h^2\sin 2\beta)-12\mu^2 h^2 }
{(3\chi h^2\sin2\beta +6 - 2h^2)^2 (3\chi^2 h^2 -2\chi h^2\sin 2\beta +4)}+V_D
\eea
where 
\be
V_D =  \frac{9(g^2+g'^2)h^4\cos^2 2\beta}{2(3\chi h^2\sin 2\beta +6-2h^2)^2}.
\ee
It is easy to see that, for fixed $\beta$ in the regime
\be
\chi h^2\gg 1 \gg h^2
\label{eq:xibig}
\ee
the potential is dominated by the $D$-term,
\be
V_D \approx \frac{g^2+g'^2}{2\chi^2}\cot^2 2\beta,
\label{VDasymp}
\ee
except very near the $D$-flat direction, more precisely, for $\sin2\beta\gg2/(\chi h^2).$
It is clear, however, that slow-roll conditions (to be reviewed below) will not be satisfied 
in these circumstances, because 
\be
\Big|\frac{1}{V_D} \frac{\pa V_D}{\pa \beta}\Big| = \frac{8}{|\sin4\beta|}
\ee
cannot be made small.  

On the other hand, in the $D$-flat direction where $\tan\beta=1$, both
$V_D$ and ${\pa V_D}/{\pa \beta}$ vanish, and ${\pa^2 V_D}/{\pa^2 \beta}>0.$  
Thus, this is a minimum stable against fluctuations in $\beta.$
Then, in the regime described in Eq.~(\ref{eq:xibig}), we have 
\be
V = -\frac{2[3 \Lambda^2 +2\mu\chi h^2 ( 2 \mu h^2 + 3 \Lambda)]}{3\chi^4h^6},
\ee
or for $\Lambda = 0$, $V = -8\mu^2/(3\chi^3 h^2)$. 
In either case, $V\to0$ through {\it negative} values, and so is unsuitable for an inflationary scenario.

\section{\nmssm}
Evidently, the failure of this \mssm{} to emulate the result of a
nonminimal coupling in the nonsupersymmetric case is  because of the
absence of a cubic term in the
superpotential, leading to the absence (in the $D$-flat direction) of an analog to the dimensionless
self-interaction  $\lambda(\phi^\dagger\phi)^2.$ (Of course inclusion of the usual 
Yukawa cubic terms in $P$ would lead to quartic terms in $V$, but we need quartic terms that 
contain only fields that occur in $X$ in order to produce a suitable flat potential). 
It  is therefore
natural to turn to the \nmssm, which clearly can provide a cure to this
behavior.  Thus we consider the superpotential 
\be
P = \lambda S H_1 H_2 + \frac{\rho}{3}S^3,
\ee
where $S$ is a gauge singlet. There are then two natural choices for $X(\phi)$, to wit 
\bea 
X &=& \chi H_1 H_2\quad \hbox{or}\label{eq:Xixih}\\
X &=& \chi S^2\label{eq:XixiS}.
\eea
Note that in the case of Eq.~(\ref{eq:Xixih}) we can choose both $\chi$ and $\lambda$ to be real. 

If we choose Eq.~(\ref{eq:Xixih}) and consider the case when $S$ is small but $h$ is large, 
\be
S=0, H_1 = (h_1, 0)^T, H_2 = (0, h_2)^T
\label{eq:sdflat}
\ee
then we find
\be
V = \frac{9(2\lambda^2 h^4\sin^2 2\beta + (g^2+g'^2)h^4\cos^2 2\beta)}
{2(3\chi h^2\sin 2\beta + 6 - 2h^2 )^2}
\label{eq:vtot}
\ee
which, taking 
the  the Eq.~(\ref{eq:xibig}) limit, becomes  
\be
V\approx \left(\frac{\lambda}{\chi}\right)^2 + \frac{g^2+g'^2}{2\chi^2}\cot^2 2\beta,
\label{Vasymp}
\ee
which is (of course) minimized for  $\tan\beta = 1$. For this value of 
$\tan\beta$, Eq.~(\ref{eq:vtot}) gives   
\be
V = \frac{9\lambda^2 h^4}{(3\chi h^2 +6 - 2 h^2 )^2},\label{Vxih}
\ee
a smoothly increasing function of $h^2$, 
precisely the behavior that we can expect to yield inflation.  
This is the supersymmetric analog of the model of Ref.~\cite{Bezrukov:2007ep}. We plot 
$V$ in Figure~1, for $\xi = 5 \times 10^4$ and $\lambda = 1$.

\EPSFIGURE[t]{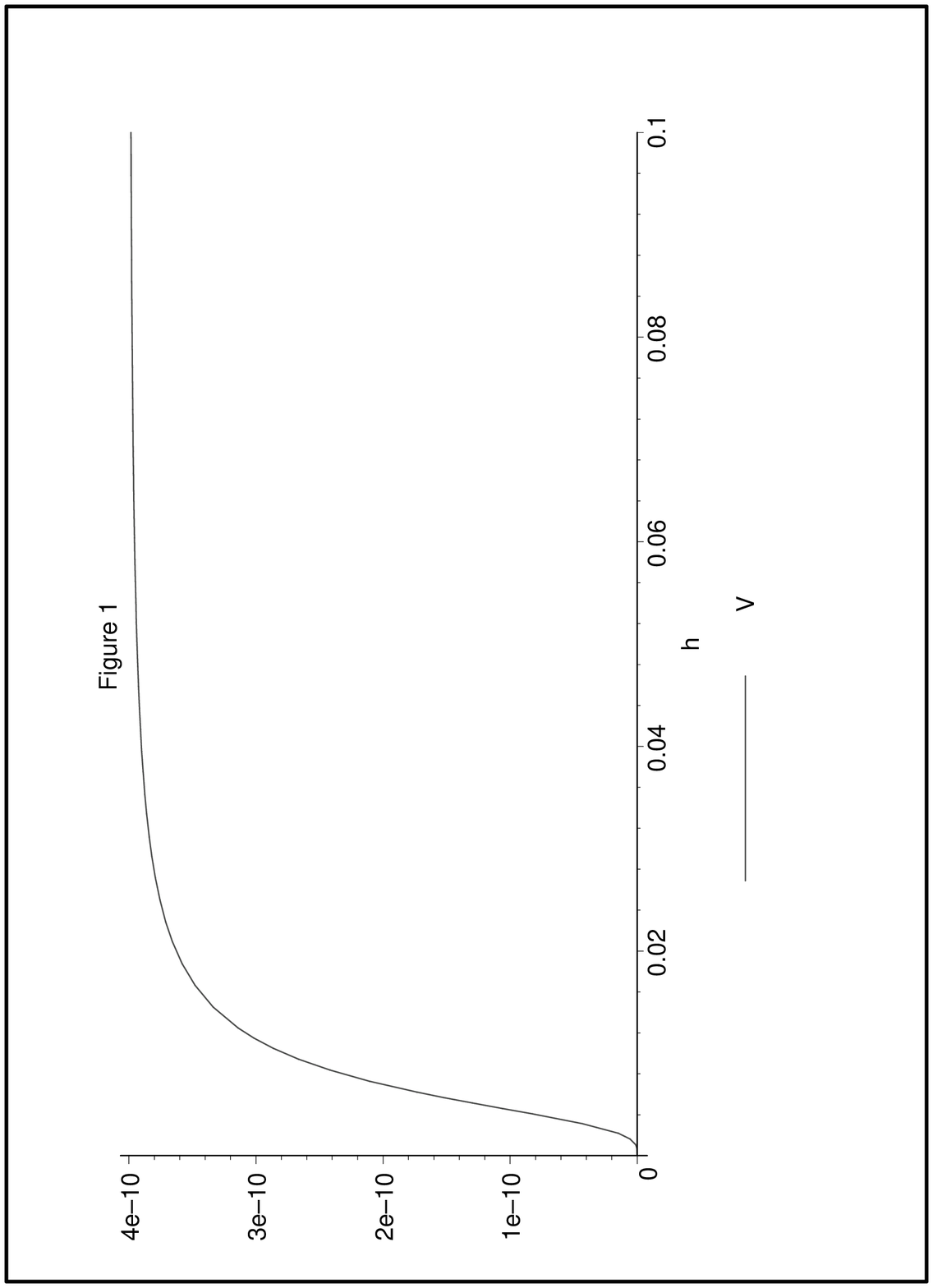,angle= -90,width=4.2in}{Plot of $V$ against $h$}

On the  other hand, if we choose Eq.~(\ref{eq:XixiS}) and $h$ small,
then we find\footnote{In principle, we cannot choose both $\rho$ and
$\chi$ real.  For our purposes, it is convenient to take $\chi$ real but
$\rho$ possibly complex.} 
\be
V = 36|\rho^2|\frac{[\chi(S^2 + S^{*2})-2]S^2 S^{*2}}
{[3\chi (S^2 + S^{*2})-2SS^{*}+6]^2[\chi (S^2 + S^{*2})-6\chi^2S S^{*}-2]}.
\label{VxiS}
\ee
Writing $S = se^{i\theta}$ we have 
\be
V = 9|\rho^2|\frac{(\chi s^2\cos 2\theta-1])s^4}
{(3\chi s^2 \cos 2\theta -s^2+3)^2(\chi s^2\cos 2\theta -3 \chi^2s^2-1]}
\label{eq:VxiSb}
\ee
which in the limit $\chi s^2 \gg 1 \gg s^2$ gives
\be
V = -\frac{|\rho^2|}{3\chi^3\cos 2\theta}.
\ee
Thus for fixed $\theta$, $V$ approaches a constant, 
but if $\chi\cos 2\theta > 0$ then this constant 
is negative. If we choose $\chi\cos 2\theta < 0$ in 
order to obtain $V > 0$ then we see from 
Eq.~(\ref{eq:VxiSb}) that as $s$ reduces, then $V$ {\it increases\/}, and we eventually 
encounter a pole in $V$, when 
\be 
\chi s^2\cos 2\theta = -1+s^2/3
\ee
Thus this case is not suitable for inflation.

\section{Slow-roll analysis}
Let us pursue the more promising case of Eq.~(\ref{Vxih}) in more detail and calculate the slow-roll 
parameters. Evidently these relate to fluctuations and so we must consider the kinetic energies 
of the relevant fields. It is easy to show using Eq.~(\ref{eq:omxi}) that in general the scalar kinetic 
energy takes the form 
\be
{\cal L}_{\rm kin} = -e(K_{\chi})_{i j^*} \pa_m \phi_i \pa^m \phi_j^*
\label{eq:kin}
\ee
where $e$ is the vierbein and
\be
K_{\chi}  = -3\ln (-\Omega_{\chi}/3).
\ee
Let us now briefly review the slow-roll paradigm
(see for example Ref.~{\cite{Baumann:2009ds}). Suppose the action that describes the inflaton $\phi$ takes the form
\be
S = \int d^4 x\, e\left[-\frac{1}{2}f(\phi)\pa^m\phi \pa_m\phi - V(\phi)\right].
\ee
Then neglecting its spatial variation, the equation of motion for  $\phi$
is
\be
f(\phi)\phiddot + 3fH\phidot + \frac{1}{2}f' (\phi) \phidot^2 + \frac{dV}{d\phi} = 0
\label{eq:Heqn}
\ee
where 
\be
H^2 = \frac{1}{3}\left(\frac{1}{2}f (\phi) \phidot^2 + V(\phi)\right).
\ee
Then the slow-roll parameter $\epsilon$ is given by
\be
\epsilon = -\frac{\Hdot}{2H^2} \approx \frac{1}{2f}\left(\frac{V'}{V}\right)^2
\label{eq:epsdef}
\ee
where the latter approximation is valid 
if $f(\phi)\phiddot$ and $f'(\phi)\phidot^2$ 
can both be safely neglected in Eq.~(\ref{eq:Heqn}). 

The number of $e$-folds before inflation ends is given by 
\be
N = \int^{\phi}_{\phi_{\rm end}}\, \frac{H}{\phidot}\, d\phi \approx 
\int^{\phi}_{\phi_{\rm end}}\, \frac{fV}{V'}\, d\phi
\label{eq:Ndef}
\ee
and the second slow-roll parameter $\eta$ is given by 
\be
\eta = \frac{V^{''}}{fV}.
\label{eq:etadef}
\ee
In the slow-roll regime, $\epsilon, |\eta| \ll 1$ and 
slow-roll ends when $\epsilon(\phi_{\rm end}) \approx 1$.

We now apply this formalism to the \nmssm{} doublet case. We first suppose that 
$\tan\beta = 1$, that is we consider the $D$-flat direction.  
From Eqs.~(\ref{eq:omxi}), (\ref{eq:dflat}), (\ref{eq:Xixih}), (\ref{eq:kin})
we find that for large $\chi$ (as defined in Eq.~(\ref{eq:xibig})) the dominant 
contribution to the inflaton kinetic energy term is 
\be
L_{\rm kin} = 3e\left(\frac{\pa_m h}{h}\right)^2.
\label{kinh}
\ee

Armed with this result we immediately obtain (given Eq.~(\ref{eq:xibig})) 
from Eqs.~(\ref{eq:epsdef}), (\ref{eq:Ndef}) and (\ref{eq:etadef})  that 
\bea
\epsilon &=& \frac{16}{3\chi^2 h^4},\\
N &=& \frac{3\chi (h^2 -h_{\rm end}^2)}{8},\\ 
\eta &=& -\frac{4}{\chi h^2}.
\eea
We see that the slow-roll conditions $\epsilon, |\eta| \ll1$ are satisfied 
given Eq.~(\ref{eq:xibig}). Moreover it is easy to verify that 
the approximations made 
in neglecting $f(\phi)\phiddot$ and $f'(\phi)\phidot^2$ 
in Eq.~(\ref{eq:Heqn}) are well justified. 
The results are similar to those of 
Ref.~\cite{Bezrukov:2008ej}. For $\epsilon (h_{\rm end}) \sim 1$, and using 
$\chi \sim 5 \times 10^4$ as in Ref.~\cite{Bezrukov:2008ej}, 
we obtain $h_{\rm end} \sim  0.7 \times 10^{-2}M_P$, in the neighborhood of 
the \gut{} scale. 

Suppose that $\tan\beta \neq 1$ at large fixed $h$. Might it be that the slow-roll 
conditions are satisfied with $\tan\beta \to 1$? In fact it is 
easy to show that 
\be 
\epsilon_{\beta} \sim \left(\frac{1}{V_D}\frac{\pa V_D}{\pa\beta}\right)^2 \sim 
\left(\frac{4(g^2 + g'^2)\cot 2\beta}{2\lambda^2\sin^2 2\beta + (g^2 + g'^2)\cos^2 2\beta}\right)^2
\ee
from which one sees that as $\beta\to\pi/4$, 
$\epsilon_{\beta} \sim [(g^2 + g'^2)(\pi-4\beta)/\lambda^2]^2$. Thus we see that, unlike in 
the \mssm{} case, models with a sufficiently slow approach to the $D$-flat direction might be entertained.

\section{Conclusions}

In conclusion, we have demonstrated that, in the presence of non-minimal 
gravitational interactions in the \nmssm{}, but not in the \mssm, there is 
contained in the Higgs sector a viable inflaton candidate, as in the \sm.   Although we
have not yet completed an analysis of the running of the couplings in
this model, we remarked earlier that the non-renormalization properties
of \susy{} and \sugra{} models make the behavior of the non-minimal couplings
different.  In particular, unlike in the nonsupersymmetric case, the
beta function for $\chi$ will therefore have a fixed point at $\chi=0.$ 
It would be interesting to implement refinements taking into account
renormalisation group  evolution described (and debated) for the \sm{} case
in  Refs.~\cite{Bezrukov:2007ep}-\cite{Okada:2009wz}, and explore the
consequences for the Higgs spectrum; in particular, perhaps in the
\mnmssm~\cite{Panagiotakopoulos:2000wp} where $\rho = 0,$ the
spectrum is  more constrained.

Given that these calculations seem to be concerned with the dynamics in
the region beyond the \gut{} scale, one question is whether the same
sort of results can be obtain within \sic\ 
\guts~\cite{Dimopoulos:1981zb}-\cite{Chamseddine:1982jx}  such as $SU(5)$ or $SO(10).$  Does inflation
end somewhere below the (reduced) Planck scale $M_P\simeq 2.4 \times
10^{18}~\GeV$  but above the \gut{} scale $M_U\simeq10^{-2} M_P$?  Is
the dynamics of inflation viable over such a relatively narrow range of
energies as this? 

The most important unanswered question is whether these calculations
make sense from the point of view of effective quantum field theory.  
As we discussed, this is a point of controversy in the literature, and
we hope to return to this issue in the future.  It is partly a question
of the range of scales over which perturbation theory can be used, and
it is partly a question of just what is the expansion parameter.  We
certainly have assumed that it makes sense to use the nonpolynomial
potential on energy scales $\sqrt{\xi} h \gtrsim M_P$ in the Einstein
frame associated with changes in the gravitational constant in the
Jordan frame, where the action has the familiar  polynomial form.  Even
if the arguments were correct that, in the \sm,  this model for
inflation is unnatural, will the situation be improved   as a result of
supersymmetry?  We leave this for future investigation.

\section*{\large Acknowledgements}
DRTJ thanks KITP (Santa Barbara) for hospitality and financial support.
This work was partially supported by the Royal Society through a collaboration grant, and by National Science Foundation
under Grant No. PHY99-07949.
%
%


\begin{thebibliography}{60}
%
\bibitem{Salopek:1988qh}
  D.~S.~Salopek, J.~R.~Bond and J.~M.~Bardeen,
  Phys.\ Rev.\  D {\bf 40}, 1753 (1989).


\bibitem{Bezrukov:2007ep}
  F.~L.~Bezrukov and
M.~Shaposhnikov,
Phys.\ Lett.\  B {\bf 659} (2008) 703
  [arXiv:0710.3755 [hep-th]].


\bibitem{Barvinsky:2008ia}
  A.~O.~Barvinsky, A.~Y.~Kamenshchik and A.~A.~Starobinsky,
  JCAP {\bf 0811}, 021 (2008)
  [arXiv:0809.2104 [hep-ph]].


\bibitem{DeSimone:2008ei}
  A.~De Simone, M.~P.~Hertzberg and F.~Wilczek,
  Phys.\ Lett.\  B {\bf 678}, 1 (2009)
  [arXiv:0812.4946 [hep-ph]].

\bibitem{Bezrukov:2008ej}
  F.~L.~Bezrukov, A.~Magnin and M.~Shaposhnikov,
  Phys.\ Lett.\  B {\bf 675} (2009) 88
  [arXiv:0812.4950 [hep-ph]].




\bibitem{Burgess:2009ea}
 C.~P.~Burgess, H.~M.~Lee and M.~Trott,
  JHEP {\bf 0909}, 103 (2009)
  [arXiv:0902.4465 [hep-ph]].

\bibitem{Barbon:2009ya}
  J.~L.~F.~Barbon and J.~R.~Espinosa,
  Phys.\ Rev.\  D {\bf 79}, 081302 (2009)
  [arXiv:0903.0355 [hep-ph]].

\bibitem{Bezrukov:2009db}
  F.~Bezrukov and M.~Shaposhnikov,
  JHEP {\bf 0907}, 089 (2009)
  [arXiv:0904.1537 [hep-ph]].


\bibitem{Barvinsky:2009fy}
 A.~O.~Barvinsky, A.~Y.~Kamenshchik, C.~Kiefer, A.~A.~Starobinsky and
C.~Steinwachs,
  JCAP {\bf 0912}, 003 (2009)
 [arXiv:0904.1698 [hep-ph]].


\bibitem{Barvinsky:2009ii}
A.~O.~Barvinsky, A.~Y.~Kamenshchik, C.~Kiefer, A.~A.~Starobinsky and
C.~F.~Steinwachs,
  arXiv:0910.1041 [hep-ph].


\bibitem{Okada:2009wz}
  N.~Okada, M.~U.~Rehman and Q.~Shafi,
  arXiv:0911.5073 [hep-ph].

\bibitem{Jack:1984vj}
  I.~Jack and H.~Osborn,
  Nucl.\ Phys.\  B {\bf 249}, 472 (1985).



\bibitem{Buchbinder:1992rb}
See, for example,  I.~L.~Buchbinder, S.~D.~Odintsov and I.~L.~Shapiro,
  ``Effective action in quantum gravity,'' {\it  Bristol, UK: IOP (1992) 413 p.}



\bibitem{Clark:2009dc}
  T.~E.~Clark, B.~Liu, S.~T.~Love and T.~ter Veldhuis,
  Phys.\ Rev.\  D {\bf 80}, 075019 (2009)
  [arXiv:0906.5595 [hep-ph]].
  

\bibitem{Lerner:2009xg}
 R.~N.~Lerner and J.~McDonald,
  arXiv:0909.0520 [hep-ph].

\bibitem{Ellwanger:2009dp}
 U.~Ellwanger, C.~Hugonie and A.~M.~Teixeira,
arXiv:0910.1785 [hep-ph].

\bibitem{Wess:1992cp}
  J.~Wess and J.~Bagger,\
{\it  Princeton, USA: Univ. Pr. (1992) 259 p}


\bibitem{Baumann:2009ds}
  D.~Baumann,
  arXiv:0907.5424 [hep-th].

\bibitem{Panagiotakopoulos:2000wp}
  C.~Panagiotakopoulos and A.~Pilaftsis,
  Phys. \ Rev. \  D  { \bf 63 } (2001) 055003 
  [arXiv:hep-ph/0008268]. 

\bibitem{Dimopoulos:1981zb}
  S.~Dimopoulos and H.~Georgi,
  Nucl.\ Phys.\  B {\bf 193}, 150 (1981).

\bibitem{Ibanez:1981yh}
  L.~E.~Ibanez and G.~G.~Ross,
  Phys.\ Lett.\  B {\bf 105}, 439 (1981).


\bibitem{Einhorn:1981sx}
  M.~B.~Einhorn and D.~R.~T.~Jones,
  Nucl.\ Phys.\  B {\bf 196}, 475 (1982).

\bibitem{Chamseddine:1982jx}
  A.~H.~Chamseddine, R.~L.~Arnowitt and P.~Nath,
  Phys.\ Rev.\ Lett.\  {\bf 49}, 970 (1982).

\end{thebibliography}
\end{document}